\documentclass[final,1p,times]{elsarticle}

 \usepackage{graphicx}

\usepackage{amssymb,amsmath}

\journal{Computer Physics Communications}

\begin{document}

\begin{frontmatter}



\title{Numerical modeling of gravitational wave sources accelerated by OpenCL}


\author[GK]{Gaurav Khanna}
\author[JM]{Justin McKennon}

\address[GK]{Physics Department, University of Massachusetts Dartmouth,\\
            North Dartmouth, MA 02747}

\address[JM]{Electrical \& Computer Engineering, University of Massachusetts Dartmouth,\\
            North Dartmouth, MA 02747}

\begin{abstract}
In this work, we make use of the OpenCL framework to accelerate an EMRI modeling application using the hardware accelerators -- Cell BE and Tesla CUDA GPU. We describe these compute technologies and our parallelization approach in detail, present our performance results, and then compare them with those from our previous implementations based on the native CUDA and Cell SDKs. The OpenCL framework allows us to execute {\it identical} source-code on both architectures and yet obtain strong performance gains that are comparable to what can be derived from the native SDKs.  
\end{abstract}


\end{frontmatter}



\section{Introduction}

In recent decades there has been a tremendous rise in numerical computer simulations in nearly every area of science and engineering. This is partly due to the development of cluster computing that involves putting together ``off-the-shelf'' computing units i.e. commodity desktop computers into a configuration that would achieve the same level of performance, or even outperform, traditional supercomputers at a fraction of the cost. The main reason behind the significant cost benefit of cluster computing is that it is entirely based on mass-produced, common desktop computers. Computational science has benefited and expanded tremendously in the last decade due to rapid improvements in CPU performance ({\em Moore's Law}) and major price drops due to mass production and intense competition. 

However, a few years ago the computer industry hit a serious {\em frequency wall}, implying that increasing the processor's clock-rate for gains in performance could not be done indefinitely, due to increases in power consumption and heat generation ({\em power wall}). This led all the major processor manufacturers toward multi-core processor designs. Today, nearly all commodity desktop and laptop processors are multi-core processors which combine two or more independent computing cores on a single chip. Thus, manufacturers continue to pack more power in a processor, even though their clock-frequencies have not risen (and have stabilized at around 3 GHz). 

It is interesting to note that there are other computing technologies that are based on a {\em many}-core design and their overall performance has continued to increase at a rate much higher than that of traditional multi-core processors. These technologies have typically been employed in desktop graphics cards (GPUs) and consumer gaming consoles. For example, {\em Compute Unified Device Architecture} (CUDA)~\cite{cuda} and {\em Cell Broadband Engine Architecture} (CBEA)~\cite{cell} are many-core architectures designed to provide high performance and scaling for multiple hardware generations. The Cell Broadband Engine (Cell BE), which is the first incarnation of the CBEA, was designed by a collaboration between Sony, Toshiba, and IBM. The Cell BE was originally intended to be used in game consoles (namely Sony's Playstation 3~\cite{ps3}) and consumer electronics devices, but the CBEA itself was not solely designed for this purpose and is been used in areas such as high-performance computing as well (IBM's Cell blades~\cite{blades}, LANL RoadRunner~\cite{roadrunner}). These many-core technologies are sometimes referred to as {\em hardware accelerators}, and have recently received a significant amount of attention from the computational science community because they can provide significant gains in the overall performance of many numerical simulations at a relatively low cost. 

However, these accelerators usually employ a rather unfamiliar and specialized programming model that often requires advanced knowledge of their hardware design. In addition, they typically have their own vendor- and design- specific software development framework, that has little in common with others: CUDA SDK for Nvidia's GPUs; ATI Stream SDK for ATI's GPUs; IBM Cell SDK for the Cell BE, while traditional multi-core processors (Intel, AMD) typically involve OpenMP-based parallel programming. All these SDKs enable parallel software development on their respective hardware, and offer programmability in the ubiquitous C programming language in conjunction with a set of libraries for memory management. Yet, the details involved in the programming are remarkably different for each such architecture. Therefore, for a computational scientist, with limited time and resources available to spend on such specialized software engineering aspects of these architectures, it becomes exceedingly difficult to embrace and make effective use of these accelerators for furthering science. 

Over the past year, under Apple's leadership, an open standard has been proposed to ``unify'' the software development for all these different computer architectures under a single standard -- the {\em Open Computing Language} (OpenCL)~\cite{opencl}. All major processor vendors (Nvidia, AMD/ATI, IBM, Intel, etc.) have adopted this standard and have just released support for OpenCL for their current hardware. In this work, we perform a careful performance-evaluation of OpenCL for scientific computing on several different hardware architectures. 

The scientific application that we will concentrate on in this work is an application from the Numerical Relativity (NR) community -- the EMRI Teukolsky Code which is a finite-difference, linear, hyperbolic, inhomogeneous partial difference equation (PDE) solver~\cite{mystuff}. It should be noted that in recent work~\cite{justin} this same application was accelerated using a Tesla GPU and Cell BE . In that work, native SDKs were used to perform this optimization i.e. CUDA SDK for the Nvidia Tesla GPU and IBM Cell SDK for the Cell BE. In our current work, we achieve the same using the OpenCL framework instead, and compare the outcome with previous work~\cite{justin}. The main advantage of our current approach is that the exact same OpenCL-based source-code executes on both accelerator hardware, therefore yielding a tremendous saving in the code development effort. It is also worth pointing out that our NR application is of a type that may also arise in various other fields of science and engineering, therefore we expect that our work would be of interest to the larger community of computational scientists. 

This article is organized as follows: In Section 2, we provide a very brief introduction to multi- and many- core processor architectures and also the OpenCL framework. In Section 3, we introduce the EMRI Teukolsky Code, the relevant background gravitational physics and the numerical method used by the code. Next, we emphasize aspects of OpenCL and the compute hardware relevant to our implementation in Section 4. In Section 5, we present the overall performance results from our OpenCL-based parallelization efforts. Finally in Section 6, we summarize our work and make some conclusive remarks.

\section{Computing Technologies}

In this section we describe in some detail the many-core compute hardware accelerators and the software development framework that are under consideration in our work.

\subsection{Multi-core \& Many-core Processor Architectures} 

As mentioned above, all processor manufacturers have moved towards multi-core designs in the quest for higher performance. At the time of the writing of this article, high-end desktop processors by Intel and AMD have a maximum of six (6) cores. On the other hand, there are other computing technologies that have been in existence for several years that have traditionally had many more compute cores than standard desktop processors. As mentioned before, these are sometimes referred to as hardware accelerators and have a many-core design. Examples of these accelerators include GPUs and the Cell BE. 

The Cell BE~\cite{cell} is a totally redesigned processor that was developed collaboratively by Sony, IBM and Toshiba primarily for multimedia applications. This processor has a general purpose (PowerPC) CPU, called the PPE (that can run two (2) software threads simultaneously) and eight (8) special-purpose compute engines, called SPEs available for raw numerical computation. Each SPE can perform vector operations, which implies that it can compute on multiple data, in a single instruction (SIMD). All these compute elements are connected to one another through a high-speed interconnect bus (EIB). Note that because of this heterogeneous design, the Cell BE is very different from traditional multi-core processors. The outcome of this distinctive design is that a single, 3.2 GHz (original 2006/2007) Cell BE has a peak performance of over 200 GFLOP/s in single-precision floating-point computation and 15 GFLOP/s in double-precision operations. It should be noted that the current (2008) release of the Cell BE, called the {\em PowerXCell}, has design improvements that bring the double-precision performance up to 100 GFLOP/s. One challenge introduced by this new design, is that the programmer has to explicitly manage the data transfer between the PPE and the SPEs. The PPE and SPEs are equipped with a DMA engine -- a mechanism that enables data transfer to and from main memory and each other. The parallel programming model on Cell BE allows for the use of SPEs for performing different tasks in a workflow ({\em task parallel} model) or performing the same task on different data ({\em data parallel} model). 

In the CUDA context, the GPU (called {\em device}) is accessible to the CPU (called {\em host}) as a co-processor with its own memory. The device executes a function (usually referred to as a {\em kernel}) in a data parallel model i.e. a number of threads run the same program on different data. The many-core architecture of the GPU makes it possible to apply a kernel to a large quantity of data in one single call. If the hardware has a large number of cores, it can process them all in parallel (for example, Nvidia's Tesla GPU has as many as 240 compute cores clocked at 1.3 GHz). In the area of high performance computing, this idea of massive parallelism is extremely important. The Tesla GPU can also perform double-precision floating point operations, at a performance comparable to that of the PowerXCell mentioned above, which happens to be an order-of-magnitude lower than its performance in single-precision. Despite that fact, a Tesla GPU's peak double-precision performance is higher than that of a typical multi-core processor, and future GPU designs promise to address this large disparity between their double and single precision performance. In addition, GPUs provide significant flexibility in terms of memory management: Six (6) main types of memory exist in the form of registers, local memory, shared memory, global memory, constant memory and texture memory. We will not attempt to go into detail with these different memory arrangements in this document; instead we will simply refer the reader to online resources on this somewhat involved topic~\cite{cuda}.

\subsection{The Open Computing Language} 

The main software framework that is under consideration in this work is the Open Computing Language. As mentioned already, OpenCL is a new framework for programming across a wide variety of computer hardware architectures (CPU, GPU, Cell BE, etc). In essence, OpenCL incorporates the changes necessary to the programming language C, that allow for parallel computing on all these different processor architectures. In addition, it establishes numerical precision requirements to provide mathematical consistency across the different hardware and vendors -- a matter that is of significant importance to the scientific computing community. Computational scientists would need to rewrite the performance intensive routines in their codes as OpenCL kernels that would be executed on the compute hardware. The OpenCL API provides the programmer various functions from locating the OpenCL enabled hardware on a system to compiling, submitting, queuing and synchronizing the compute kernels on the hardware. Finally, it is the OpenCL runtime that actually executes the kernels and manages the needed data transfers in an efficient manner. As mentioned already, most vendors have released an OpenCL implementation for their own hardware. 

From a programming standpoint, OpenCL is a relatively simple language to use, provided that the programmer has a strong C language background. Operations are performed with respect to a given context -- the creation of which is the first step to initializing and using OpenCL. Each context can have a number of associated devices (CPU, GPU) and within a context, OpenCL guarantees a relaxed memory consistency between devices. OpenCL uses buffers (1 dimensional blocks of memory) and images (2 and 3 dimensional blocks of memory) to store the data of the kernel that is to be run on the specified device. Once memory for the kernel data has been allocated and a device has been specified, the kernel program (the program which the programmer intends to run on the device) needs to be loaded and built. To call a kernel, the programmer must build a kernel object. Once the kernel object has been built and the arguments to the kernel have been set, the programmer must create a command queue. All of the computations done on the device are done using a command queue. The command queue is essentially a virtual interface for the device and each command queue has a one-to-one mapping with the device. Once the command queue has been created, the kernel can then be queued for execution. The total number of elements or indexes in the launch domain is referred to as the {\em global work size} and individual elements are referred to as {\em work items}. These work items can be combined into {\em work groups} when communication between work items is required. The kernel can be executed on a 1, 2 or 3 dimensional domain of indexes -- all of which execute in parallel, given proper resources. 

\section{EMRI Teukolsky Code}

In our earlier work~\cite{justin} we describe the EMRI Teukolsky Code in detail, and also present the relevant background gravitational physics. Therefore, we simply reproduce the relevant section from Ref.~\cite{justin} below for completeness with minimal alterations. 

Many gravitational wave observatories~\cite{ligo} are currently being built all over the globe. These laboratories will open a new window into the Universe by enabling scientists to make astronomical observations using a completely new medium -- gravitational waves (GWs) as opposed to electromagnetic waves (light). These GWs were predicted by Einstein's relativity theory, but have not been directly observed because the required experimental accuracy was simply not advanced enough (until very recently). 

Numerical Relativity is an area of gravitational physics that is focussed on the numerical modeling of strong sources of GWs -- collisions of compact astrophysical objects such as neutron stars and black holes. Thus, it plays an extremely important role in this new and upcoming area of GW astronomy. The specific NR application that we consider in this paper is one that evolves the GWs generated by a compact object (such as a star of the size of our own Sun) that has a decaying orbit around a supermassive black hole. Such large black holes -- often more massive than a million times our Sun -- lurk at the center of most galaxies and routinely devour smaller stars and black holes. Such processes are commonly referred to as extreme mass-ratio inspirals (EMRIs) in the relevant literature. The low-frequency gravitational waves emitted from such EMRI systems are expected to be in good sensitivity band for the upcoming space-borne gravitational wave detectors -- such as the ESA/NASA Laser Interferometer Space Antenna (LISA) mission~\cite{lisa}. Studies of the dynamics and the orbital evolution of a binary system in the extreme mass-ratio limit is therefore an important issue for low-frequency gravitational wave detection. 

Because of the extreme mass-ratio, the small object orbiting around the central supermassive black hole can be modeled as a small structure-less object, and the problem can be addressed within black hole perturbation theory. This is where the Teukolsky equation becomes relevant. This equation governs the evolution of the perturbations of rotating (Kerr) black holes, with the small object acting as a ``source'' of the perturbations. In other words, the Teukolsky equation is essentially a linear wave equation in Kerr space-time geometry, with the small object acting as generator of the gravitational waves. Thus, to numerically model an EMRI scenario, we solve the inhomogeneous Teukolsky equation in the time-domain.

The next two subsections provide more detailed information on this equation and the associated numerical solver code.

\subsection{Teukolsky Equation}

The Teukolsky master equation describes scalar, vector and tensor field perturbations in the space-time of Kerr black holes~\cite{teuk}. In Boyer-Lindquist coordinates, this equation takes the form
\begin{eqnarray}
\label{teuk0}
&&
-\left[\frac{(r^2 + a^2)^2 }{\Delta}-a^2\sin^2\theta\right]
         \partial_{tt}\Psi
-\frac{4 M a r}{\Delta}
         \partial_{t\phi}\Psi \nonumber \\
&&- 2s\left[r-\frac{M(r^2-a^2)}{\Delta}+ia\cos\theta\right]
         \partial_t\Psi\nonumber\\  
&&
+\,\Delta^{-s}\partial_r\left(\Delta^{s+1}\partial_r\Psi\right)
+\frac{1}{\sin\theta}\partial_\theta
\left(\sin\theta\partial_\theta\Psi\right)+\nonumber\\
&& \left[\frac{1}{\sin^2\theta}-\frac{a^2}{\Delta}\right] 
\partial_{\phi\phi}\Psi +\, 2s \left[\frac{a (r-M)}{\Delta} 
+ \frac{i \cos\theta}{\sin^2\theta}\right] \partial_\phi\Psi  \nonumber\\
&&- \left(s^2 \cot^2\theta - s \right) \Psi = -4\pi (r^2 + a^2 \cos^2 \theta)\, T ,
\end{eqnarray}
where $M$ is the mass of the black hole, $a$ its angular momentum per unit mass, $\Delta = r^2 - 2 M r + a^2$ and $s$ is the ``spin weight'' of the field. The $s = \pm 2$ versions of these equations describe the radiative degrees of freedom of the gravitational field, and thus are the equations of interest here. As mentioned previously, this equation is an example of linear, hyperbolic, inhomogeneous PDEs that arise in several areas of science and engineering, and can be solved numerically using a variety of finite-difference schemes. The quantity $T$ in Eq. (1) is the ``source'' term as mentioned in the previous section. It plays an extremely critical role in this work and that will be discussed in detail, later in this paper. Ref.~\cite{mystuff} has a mathematical formula for this quantity and to save space, we will not reproduce that expression here.

\subsection{Numerical Method}

Ref.~\cite{krivan} demonstrated stable numerical evolution of Eq.\ (\ref{teuk0}) for $s=-2$ using the well-known Lax-Wendroff numerical evolution scheme. Our Teukolsky Code uses the exact same approach, therefore the contents of this section are largely a review of the work presented in the relevant literature~\cite{mystuff}. 

Our code uses the tortoise coordinate $r^*$ in the radial direction and azimuthal coordinate $\tilde{\phi}$. These coordinates are related to the usual Boyer-Lindquist coordinates by
\begin{eqnarray}
dr^* &=& \frac{r^2+a^2}{\Delta}dr 
\end{eqnarray}
and
\begin{eqnarray}
d\tilde{\phi} &=& d\phi + \frac{a}{\Delta}dr \; . 
\end{eqnarray}  
Following Ref.~\cite{krivan}, we factor out the azimuthal dependence and use the ansatz,
\begin{eqnarray}
\label{eq:psiphi}
\Psi(t,r^*,\theta,\tilde{\phi}) &=& e^{im\tilde{\phi}} r^3 \Phi(t,r^*,\theta) .
\end{eqnarray}
Defining
\begin{eqnarray}
\Pi &\equiv& \partial_t{\Phi} + b \, \partial_{r^*}\Phi \; , \\
b & \equiv &
\frac { {r}^{2}+{a}^{2}}
      { \Sigma} \; , 
\end{eqnarray}
and
\begin{eqnarray}
\Sigma^2 &\equiv &  (r^2+a^2)^2-a^2\,\Delta\,\sin^2\theta
\; 
\label{pi_eq}
\end{eqnarray} 
allows the Teukolsky equation to be rewritten as
\begin{eqnarray}
\label{eq:evln}
\partial_t \mbox{\boldmath{$u$}} + \mbox{\boldmath{$M$}} \partial_{r*}\mbox{\boldmath{$u$}} 
+ \mbox{\boldmath{$Lu$}} + \mbox{\boldmath{$Au$}} =  \mbox{\boldmath{$T$}} ,
\end{eqnarray}
where 
\begin{equation}
\mbox{\boldmath{$u$}}\equiv\{\Phi_R,\Phi_I,\Pi_R,\Pi_I\}
\end{equation}
is the solution vector. The subscripts $R$ and $I$ refer to the real and imaginary parts respectively (note that the Teukolsky function $\Psi$ is a complex valued quantity). Explicit forms for the matrices {\boldmath{$M$}}, {\boldmath{$A$}} and {\boldmath{$L$}} can be easily found in the relevant literature~\cite{krivan}. Rewriting Eq.\ (\ref{eq:evln}) as 
\begin{equation}
\partial_t \mbox{\boldmath{$u$}} + \mbox{\boldmath{$D$}}
\partial_{r^*} \mbox{\boldmath{$u$}}
=  \mbox{\boldmath{$S$}}\; , 
\label{new_teu2}
\end{equation}
where
\begin{equation}
 \mbox{\boldmath{$D$}} \equiv \left(\begin{matrix}
                    b &   0   &  0  &  0 \cr
                    0  &   b   &  0  &  0 \cr
                    0  &   0   &  -b  &  0 \cr
                    0  &   0   &  0  &  -b \cr
                \end{matrix}\right),
\label{d_matrix}
\end{equation}
\begin{equation}
\mbox{\boldmath{$S$}} =\mbox{\boldmath{$T$}} -(\mbox{\boldmath{$M$}} - \mbox{\boldmath{$D$}})
\partial_{r^*}\mbox{\boldmath{$u$}}
- \mbox{\boldmath{$L$}}\mbox{\boldmath{$u$}} 
- \mbox{\boldmath{$A$}}\mbox{\boldmath{$u$}},
\end{equation}
and using the Lax-Wendroff iterative scheme, we obtain stable evolutions. Each iteration consists of two steps: In the first step, the solution vector between grid points is obtained from
\begin{eqnarray}
\label{lw1}
\mbox{\boldmath{$u$}}^{n+1/2}_{i+1/2} &=& 
\frac{1}{2} \left( \mbox{\boldmath{$u$}}^{n}_{i+1}
                  +\mbox{\boldmath{$u$}}^{n}_{i}\right)
- \\
&  &\frac{\delta t}{2}\,\left[\frac{1}{\delta r^*} \mbox{\boldmath{$D$}}^{n}_{i+1/2}
  \left(\mbox{\boldmath{$u$}}^{n}_{i+1}
                  -\mbox{\boldmath{$u$}}^{n}_{i}\right)
- \mbox{\boldmath{$S$}}^{n}_{i+1/2} \right] \; .\nonumber
\end{eqnarray}
This is used to compute the solution vector at the next time step,
\begin{equation}
\mbox{\boldmath{$u$}}^{n+1}_{i} = 
\mbox{\boldmath{$u$}}^{n}_{i}
- \delta t\, \left[\frac{1}{\delta r^*} \mbox{\boldmath{$D$}}^{n+1/2}_{i}
  \left(\mbox{\boldmath{$u$}}^{n+1/2}_{i+1/2}
                  -\mbox{\boldmath{$u$}}^{n+1/2}_{i-1/2}\right) \\
- \mbox{\boldmath{$S$}}^{n+1/2}_{i} \right] \, .
\label{lw2}
\end{equation}
The angular subscripts are dropped in the above equation for clarity. All angular derivatives are computed using second-order, centered finite difference expressions. 

Following Ref.~\cite{krivan}, we set $\Phi$ and $\Pi$ to zero on the inner and outer radial boundaries. Symmetries of the spheroidal harmonics are used to determine the angular boundary conditions: For even $|m|$ modes, we have $\partial_\theta\Phi =0$ at $\theta = 0,\pi$ while $\Phi =0$ at $\theta = 0,\pi$ for modes of odd $|m|$.

\begin{figure}
\centering
\includegraphics[width=5in]{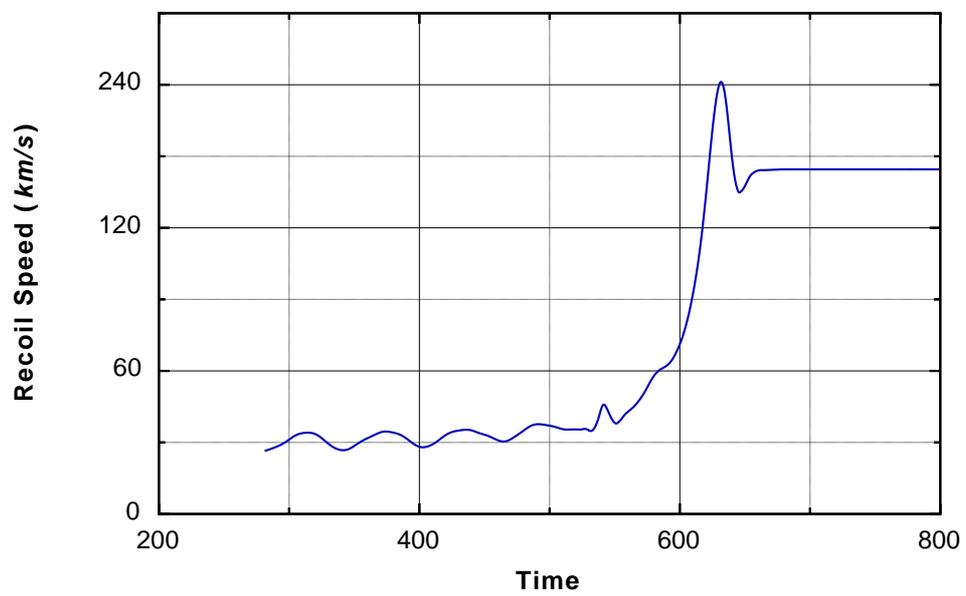}
\caption{Recoil velocity or ``kick'' experienced by an EMRI system, as a function of time as computed by our EMRI Teukolsky Code. This recoil is caused due the linear momentum carried away by the gravitational waves emitted by this collapsing system.}\label{kick}
\end{figure}

Numerical evolutions performed using the EMRI Teukolsky Code can provide a wealth of information, not only about the gravitational wave emitted by the EMRI system, but also about the behavior of the black hole binary system itself. For example, gravitational wave emission can cause the binary system to experience a {\it recoil}, much like the ``kick'' experienced upon firing a rifle. This is due to the fact that these waves not only carry away energy from the binary system, they also carry linear and angular momentum away. These recoil velocities can be in the thousands of {\it km/s} and may be responsible for ejecting the binary system from the host galaxy! In Fig.~\ref{kick} we depict a recoil experienced by a binary system with a mass-ratio of $1/10$. The larger black hole here is a rotating (Kerr) black hole with Kerr parameter $a/M = 0.5$. 

\section{OpenCL Parallel Implementation}

In our recent work~\cite{justin} the EMRI Teukolsky Code has been developed for optimized execution on the Cell/GPU hardware. That accelerated code yields a speed up of well more than an {\em order-of-magnitude} over a PPE/CPU-only code~\cite{justin}. It should be noted that the context of this computation is double-precision floating point accuracy. In single-precision this speed up would be significantly higher. The code development performed in~\cite{justin} uses Nvidia's CUDA SDK for the GPU hardware and IBM Cell SDK for the Cell BE. It is worth pointing out that in spite of the fact that the approach taken towards parallelization is identical, the actual code involving these two different architectures has little in common. This is because of the vast differences between CUDA and Cell SDK -- from the explicit memory and thread management to even the compilation process. OpenCL is uniquely positioned to address this serious problem and allows a computational scientist to experiment with different accelerator hardware without such a significant redundant software development effort. Below we describe our approach toward parallelization of the EMRI Teukolsky Code in the OpenCL framework. 

The 8 SPEs of the Cell BE and the 240 cores of the Tesla GPU are the main compute engines of these accelerator devices respectively, therefore one would want these to execute the most compute intensive tasks of a code in a data-parallel fashion. Upon performing a basic profiling of our code using the GNU profiler {\bf gprof}, we learn that simply computing the source-term $T$ (see Section 3.1) takes {\bf 99\%} of the application's overall runtime. Thus, it is natural to consider accelerating this $T$ calculation using data parallelization on the SPEs of the Cell BE and the cores of the Tesla GPU. 

A data parallel model is straightforward to implement in a code like ours. We simply perform a domain-decomposition of our finite-difference numerical grid and allocate the different parts of the grid to different SPEs or GPU cores. Essentially, each compute thread computes $T$ for a single pair of $r^*$ and $\theta$ grid values, which results in the hardware executing a few million threads in total. Note that all these calculations are independent, i.e. no communication is necessary between the SPEs/GPU threads. 

In addition, it is necessary to establish the appropriate data communication between the SPEs/GPU cores and the remaining code that is executing on the PPE/CPU respectively. We simply use {\bf clEnqueueReadBuffer, clEnqueueWriteBuffer} instructions to achieve this. Only a rather small amount of data is required to be transferred back and forth from the SPEs/GPU at every time-step of the numerical evolution. To be more specific, approximately 10 floating-point numbers are required by the SPEs/GPU to begin the computation for a specific $(r^{*},\theta)$ and then they release only 2 floating-point values as the result of the source-term $T$ computation. Because of the rather modest amount of data-transfer involved, we do not make use of any advanced memory management features on both architectures. In particular, we do not make use ``double buffering'' on the Cell BE, and we only use global memory on the GPU. 

The source-term computation in itself is rather complicated -- the explicit mathematical expression for $T$ is too long to list here. A compact expression of its form can be found in Ref.~\cite{mystuff} although that is perhaps of limited usefulness from the point of view of judging its computational complexity. A somewhat expanded version of the expression is available on slide 10 of Pullin's seminar~\cite{pullin} in the 4th CAPRA meeting (2001) at Albert-Einstein-Institute in Golm, Germany. It will suffice here to say that it is essentially a very long mathematical formula that is implemented using numerical code (approximately 2500 lines generated by computer algebra software, {\em Maple}) that uses elementary floating-point operations (no looping and very few transcendental function calls) and approximately 4000 temporary variables of the {\bf double}\footnote{It is worth pointing out that in our earlier work~\cite{justin} made use of the complex datatype in the source-term calculation, as is required for full generality. Because of the unavailability of the complex datatype in OpenCL, in  our current work we restricted our computation to cases wherein the complex datatype is unnecessary. More specifically, instead of modeling rotating (Kerr) central black holes, we only model non-rotating (Schwarzschild) black holes in our present work.} datatype. These temporary variables reside in the local memory of the SPEs/GPU cores. The basic structure of the OpenCL kernel is depicted below. The code makes use of {\it double-precision} floating-point accuracy because that is the common practice in the NR community and also a necessity for such finite-difference based evolutions, especially if a large number of time-steps are involved. We do not perform any low-level optimizations by hand (such as making use of vector operations on the CPU, SPEs etc.) on any architecture, instead we rely on mature compilers to perform such optimizations automatically. However, due to the absence of any loop structure in our source-term code, the compilers primarily make use of scalar operations to perform the computations. 

\begin{verbatim}

#pragma OPENCL EXTENSION cl_khr_fp64: enable

__kernel void 
add(__global double *thd, __global double *rd, __global double *tmp, 
                __global double *tred, __global double *timd)
{
         int gid = get_global_id(0);   /* thread's identification */

/* .. approx 4000 temporary variables as needed are declared here .. */

          r = rd[gid];               /* r* coordinate grid values */
          th = thd[gid];             /* theta coordinate grid values */

/* .. approximately 2500 lines of Maple generated code 
      for the source-term expression that uses all the 
      variables above is included here .. */

          tred[gid] =  ..              /* output variables assigned */
          timd[gid] =  ..              /* the computed values above */
  
}

\end{verbatim}

The parallel implementation outlined above is straightforward to implement in OpenCL, mainly because we have a CUDA based version of the same code to begin with~\cite{justin}. OpenCL and CUDA are very similar in style and capability, therefore developing code in these frameworks is a near identical process -- we anticipate that one could develop codes for both simultaneously with only a little extra investment of time and resources (approximately 10 -- 15\%). Resulting codes would resemble each other closely and be structured very similarly and be essentially of the same length. As mentioned before, Cell SDK is quite different and the Cell BE version of our code involves very different programming details (for example, the use of mailboxes for synchronization and using DMA calls for data exchange between PPE and SPEs). One major difference between CUDA and OpenCL worth pointing out is that OpenCL kernels currently do not support any C++ features (for example, {\em operator overloading} etc.) which can be an issue (this is the reason why we are unable to define complex number datatypes and operations in our OpenCL kernel). Code development details aside, another aspect of working with OpenCL which is perhaps somewhat challenging currently is discovering and finding workarounds for issues that appear on various different hardware platforms. On most systems, OpenCL is a {\em beta} release, therefore the specification is supported to varying degrees by different vendors. However, this is an issue that will automatically be resolved with time, as OpenCL device drivers mature and the specification is fully supported on all hardware platforms. 

\section{Performance Results}

In this section of this article, we report on the performance results from our OpenCL implementation, and also how they compare with those from our previous implementations based on CUDA and Cell SDKs. We use the following hardware for our performance tests: IBM QS22 blade system, with two (2) PowerXCell processors clocked at 3.2 GHz. This system is equipped with 16 GBs of main memory. In the GPU context, our system supports the Nvidia C1060 Tesla CUDA GPU. This system has an AMD 2.5 GHz Phenom (9850 quad-core) processor as its main CPU and four (4) GBs of memory. All these systems are running Fedora Linux as the primary operating system. Standard open-source GCC compiler suite for code development is available on all these systems. However, on the QS22 blade system we make use of the commercial IBM XLC/C++ compiler suite. We use vendor (IBM, Nvidia) supplied OpenCL libraries and compilers on both systems -- these are presently in {\it beta} development stage and are therefore likely to improve significantly in the future.

\subsection{OpenCL EMRI Teukolsky Code Performance on Cell BE Hardware}

\begin{figure}
\centering
\includegraphics[width=5in]{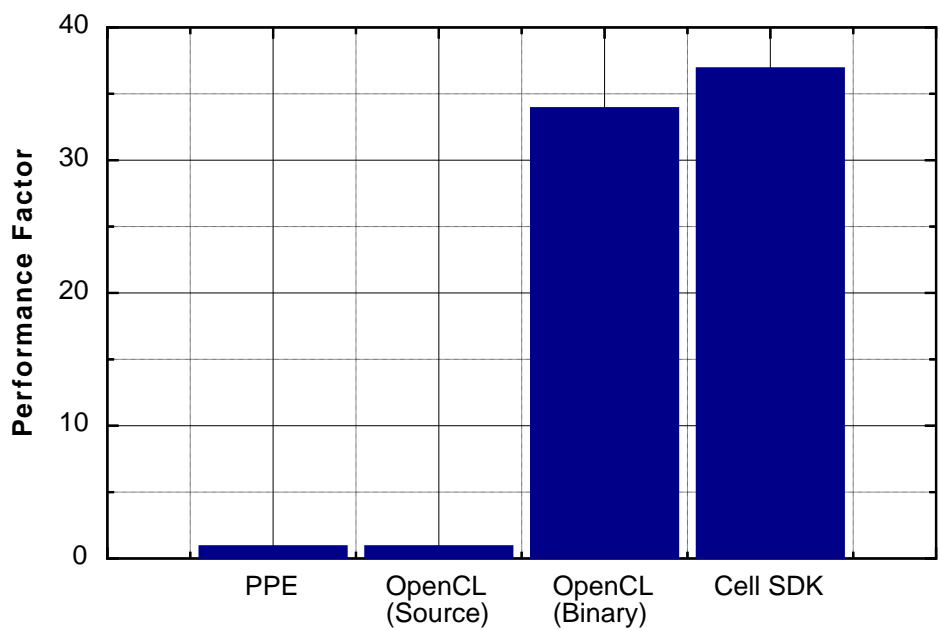}
\caption{Overall performance of the EMRI Teukolsky Code accelerated by the Cell Broadband Engine using the OpenCL framework. The baseline here is the Cell's PPE.}\label{cell}
\end{figure} 

In Fig.~\ref{cell} we show the overall performance results from our EMRI Teukolsky Code as accelerated by the Cell BE. We choose the PPE as the baseline for this comparison. Both our OpenCL (Binary) and Cell SDK based codes deliver an impressive {\bf 30x} gain in performance over the PPE. For this comparison, we use the maximum allowed {\bf local\_work\_size} of 256 in the OpenCL code. There are two remarks worth making in the context of this comparison. Firstly, the OpenCL-based performance we mention for the comparison above, results when the OpenCL kernel is {\em pre-compiled}. If the kernel is left as source-code, the kernel compilation itself strongly dominates the total runtime of the code, resulting in negligible performance gain over the PPE. These results are labelled in Fig.~\ref{cell} as OpenCL (Source). Secondly, OpenCL makes use of all the available SPEs in the QS22 blade i.e. 16 SPEs in total. Therefore, to estimate the gain from a single Cell BE processor (8 SPEs), we halve the performance gain obtained from the entire QS22 blade. 

On the Cell BE, an OpenCL based implementation delivers comparable performance to that based on the native Cell SDK. These performance gains are well over an {\em order-of-magnitude} and therefore the Cell BE with OpenCL framework has great potential for significantly accelerating many scientific applications. 

\subsection{OpenCL EMRI Teukolsky Code Performance on Tesla GPU Hardware}

\begin{figure}
\centering
\includegraphics[width=5in]{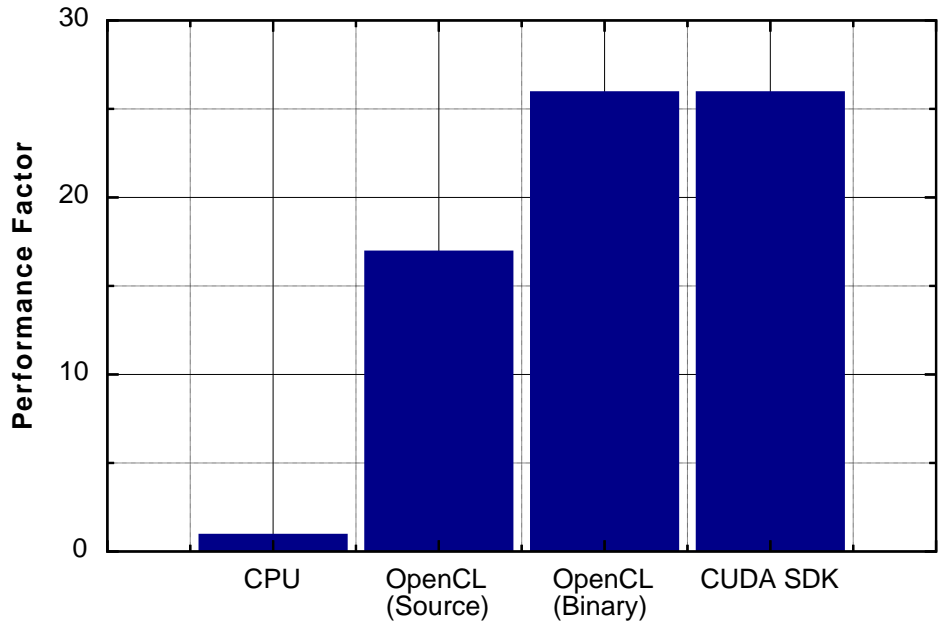}
\caption{Overall performance of the EMRI Teukolsky Code accelerated by the Tesla CUDA GPU using OpenCL. The baseline here is the supporting system's CPU -- an AMD Phenom 2.5 GHz processor.}\label{gpu}
\end{figure} 

In Fig.~\ref{gpu} we show the overall performance results from our EMRI Teukolsky Code as accelerated by the Nvidia Tesla CUDA GPU. Here we choose the CPU of the supporting system as the baseline. This CPU is a four (4) core AMD Phenom 2.5 GHz processor. We choose the baseline for these comparisons to be the single-core~\footnote{The OpenCL Teukolsky Code cannot be executed on this multi-core CPU platform because (at the time this work was conducted) AMD's OpenCL implementation does not include support for double-precision floating point operations.} performance of our EMRI Teukolsky Code on an AMD 2.5 GHz Phenom processor. 

Once again we note that the OpenCL based implementation performs comparably well to the one based on CUDA SDK ({\bf 25x} gain). These performance gains are well over an {\em order-of-magnitude} and therefore the Tesla GPU with OpenCL has great potential for significantly accelerating many scientific applications. In addition, even for the case in which the OpenCL kernel is not pre-compiled, the overall performance gain is significant. For this comparison, we use a {\bf local\_work\_size} of 128 in the OpenCL code, because a larger value yields incorrect results (it is advisable to test the robustness of the results generated, by varying the value of the {\bf local\_work\_size}).   

\subsection{Relative Performance}

\begin{figure}
\centering
\includegraphics[width=5in]{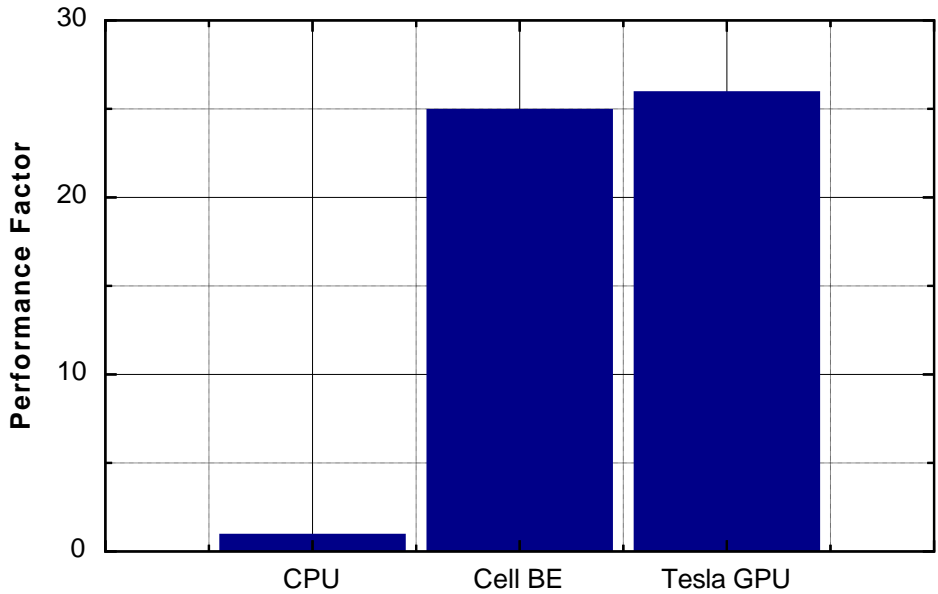}
\caption{Relative performance of the OpenCL-based EMRI Teukolsky Code on all discussed architectures -- CPU, CBE and GPU. The baseline here is the system CPU -- an AMD Phenom 2.5 GHz processor.}\label{comp}
\end{figure} 

In Fig.~\ref{comp} we depict the relative performance of all these architectures together, with the single-core 2.5 GHz AMD Phenom processor as the baseline. The results presented there are self explanatory. It is worth reporting that our estimate of the actual performance numbers in GFLOP/s obtained by our code on the single-core AMD Phenom CPU is approximately 1.0 GFLOP/s. This suggests that our code achieves 25 -- 30\% percent of peak performance on the Cell BE and Tesla GPU. Recall that due to the difficulty of using vector operations in our computations, this performance is almost entirely from scalar operations -- which suggests that our code is quite efficient. Finally, it is also worth commenting on the comparative cost associated to procuring these different hardware architectures. The 2.5 GHz AMD Phenom is very inexpensive, and a system similar to the one we used for testing can be easily obtained for under \$1,000. The Cell BE, that exhibits very impressive performance in our tests is the most expensive hardware to obtain, with its listed price on IBM's website being \$5,000 per processor. However, IBM is currently heavily discounting QS22 blades, and thus the ``street'' price is closer to \$2,500 per processor. Finally, the Nvidia Tesla GPU is currently available for approximately \$1,500. 

\section{Conclusions}

The main goal of this work is to evaluate an emerging computational platform, OpenCL, for scientific computation. OpenCL is potentially extremely important for all computational scientists because it is hardware and vendor neutral and yet (as our results suggest) able to deliver strong performance i.e. it provides {\em portability} without sacrificing {\em performance}. In this work, we consider all major types of compute hardware (CPU, GPU and even a hybrid architecture i.e. Cell BE) and provide comparative performance results based on a specific research code. 

More specifically, we take an important NR application -- the EMRI Teukolsky Code -- and perform a low-level parallelization of its most computationally intensive part using the OpenCL framework, for optimized execution on the Cell BE and Tesla CUDA GPU. We describe the parallelization approach taken and also the relevant important aspects of the considered compute hardware in some detail. In addition, we compare the performance gains we obtain from our OpenCL implementation to the gains from native Cell and CUDA SDK based implementations. 

The final outcome of our work is very similar on these architectures -- we obtain well over an {\em order-of-magnitude} gain in overall application performance. Our results also suggest that an OpenCL-based implementation delivers comparable performance to that based on a native SDK on both types of accelerator hardware. Moreover, the OpenCL source-code is {\it identical} for both these hardware platforms, which is a non-trivial benefit -- it promises tremendous savings in parallel code-development and optimization efforts. 

\section{Acknowledgements}

The authors would like to thank Glenn Volkema and Rakesh Ginjupalli for their assistance with this work throughout, many helpful discussions and also for providing useful feedback on this manuscript. GK would like to acknowledge research support from the National Science Foundation (NSF grant numbers: PHY-0831631, PHY-0902026), Apple, IBM, Sony and Nvidia. JM is grateful for support from the Massachusetts Space Grant Consortium and the NSF CSUMS program of the Mathematics Department. 





\bibliographystyle{elsarticle-num}


\end{document}